\documentstyle[12pt,preprint,aps]{revtex}

\def\thalf{{\textstyle{\frac{1}{2}}}}

\title{Interpretation of the First Data on Central Au+Au Collisions at 
$\sqrt{s}$ = 56 and 130 A GeV}
\author{
Sangyong Jeon\footnote{jeon@nta2.lbl.gov}}
\address{
{ Nuclear Science Division, Lawrence Berkeley National Laboratory,
Berkeley, CA 94720}}
\author{
Joseph Kapusta\footnote{kapusta@physics.spa.umn.edu}}
\address{
{School of Physics and Astronomy,
University of Minnesota,
Minneapolis, MN 55455}}

\begin{document}

\maketitle

\begin{abstract}
We compare three semi-microscopic theories to the first data on particle 
production in central Au+Au collisions taken at RHIC by the PHOBOS
collaboration as well as to existing data on central Pb+Pb collisions taken 
at the SPS by the NA49 collaboration.   LEXUS represents the SPS data quite 
well but not the RHIC data, whereas the wounded nucleon model does the 
opposite.  The collective tube model fails to describe any of the data.
This suggests a transition in the dynamics of particle production between 
$\sqrt{s}$ = 17 and 56 A GeV as one goes from the SPS to RHIC. 
\end{abstract}


The first data from the Relativistic Heavy Ion Collider (RHIC) at Brookhaven 
National Laboratory has been presented by the PHOBOS collaboration 
\cite{phobos}.  Their result is that the numbers of electrically charged 
hadrons per unit of pseudo-rapidity, $dN_{ch}/d\eta$, produced in the 
6\% most central Au+Au collisions at $\sqrt{s}$ = 56 and 130 A GeV and 
averaged over the 
interval $|\eta| < 1$, are 408$\pm$12(stat)$\pm$30(syst) and 
555$\pm$12(stat)$\pm$35(syst), respectively.  Pseudo-rapidity is defined as 
$\eta = \thalf \ln\left[(1+\cos\theta)/(1-\cos\theta)\right]$, where the angle 
$\theta$ is measured with respect to the beam axis. 
The previous maximum energy for heavy ion collisions 
was $\sqrt{s}$ = 17 A GeV for Pb+Pb collisions at the SPS at CERN.  
Particle production in a high energy heavy ion collision is one of the 
fundamental observables.  In this paper we report on a comparison of three
semi-microscopic theories with both the RHIC and the SPS data in an attempt to 
understand the basic dynamics of these collisions.  These theories are (1) a 
Linear EXtrapolation of Ultrarelativistic nucleon-nucleon Scattering to
nucleus-nucleus collisions (LEXUS) \cite{lexus}, (2) the Wounded Nucleon Model 
(WNM) \cite{wnm}, and (3) the Collective Tube Model (CTM) \cite{ctm}.  
We refer to these as semi-microscopic theories because they are based on input
from nucleon-nucleon collisions but are not computed with QCD.  Below we 
briefly describe these theories; for details the reader should consult the 
original papers.

The LEXUS assumes that the nucleons follow straight-line trajectories, 
striking nucleons from the other nucleus that lie in their path and 
interacting with them 
exactly as in free space.  Hadrons are produced in every nucleon-nucleon 
collision according to the parametrization
\begin{equation}
N_{ch}^{NN}(s) = 1.568 \left(\sqrt{s}-M_{min}\right)^{3/4}/s^{1/8}
\sim 1.568 s^{1/4}
\end{equation}
with $\sqrt{s}$ being the center-of-mass energy in that nucleon-nucleon 
collision and is measured in GeV.  With $M_{min} = 2m_N+m_{\pi}$ this simple 
function represents particle production in nucleon-nucleon collisions up to 
$\sqrt{s}$ = 62 GeV, excluding single diffractive events, as well as
proton-antiproton collisions at 200 GeV.  It is also known that in
nucleon-nucleon collisions the hadrons are produced with a Gaussian rapidity 
distribution centered at mid-rapidity and with a dispersion 
given by the formula
\begin{equation}
D^2_{NN}(s) = \ln\left(\frac{\sqrt{s}}{2m_N}\right) \, .
\end{equation}
Unlike pseudo-rapidity, rapidity requires knowledge of the mass of the 
particle and is defined as $y = \thalf \ln\left[(E+p_z)/(E-
p_z)\right]$, where $E$ is the energy and $p_z$ is the momentum along 
the beam axis.  When the mass goes to zero $\eta$ and $y$ coincide; for 
pions their difference is typically very small.  As a nucleon cascades 
through the other nucleus it loses energy, and this is taken into 
account via an evolution equation which is solved numerically.  All 
parameters in LEXUS are fit to nucleon-nucleon data and nothing should 
be adjusted to fit nucleus-nucleus data.  The above information is 
folded together with a constant inelastic nucleon-nucleon cross section 
$\sigma_{inel}$ and with a realistic density distribution for the 
colliding nuclei.

The WNM defines a nucleon to be wounded the first time it undergoes 
an inelastic collision with a nucleon from the other nucleus.  A 
wounded nucleon is assumed to produce 1/2 of the average charged hadrons in 
a nucleon-nucleon collision at the same energy.  Once it is wounded it 
cannot produce any more, although it can 
strike an unwounded nucleon and that one can produce particles.  The total 
number of charged hadrons produced by $n_P$ wounded projectile nucleons and 
$n_T$ wounded target nucleons is
\begin{equation}
N_{ch}(n_P,n_T) = \frac{n_P + n_T}{2} N_{ch}^{NN}(s) \, .
\end{equation}
These hadrons are assumed to be distributed in rapidity in a Gaussian way, 
centered at the nucleon-nucleon rest frame, and with a dispersion given by eq. 
(2).  There is no energy loss assigned to the nucleons as they 
strike and wound other nucleons.  Otherwise the geometrical folding to compute 
the number of wounded nucleons is standard and is done in exactly the same way 
as LEXUS.

The CTM describes a nucleus-nucleus collision as a set of aligned tube-tube 
collisions.  One tube is taken from the projectile nucleus and one from the 
target.  The cross sectional area of the tubes is $\sigma_{inel}$.  If 
one tube contains $n_P$ and the other tube $n_T$ participants then the 
center-of-mass energy available for particle production is
\begin{equation}
s(n_P,n_T) = 4n_Pn_T p_{cm}^2 + (n_P+n_T)^2 m_N^2 \, ,
\end{equation}
where $p_{cm}$ is the beam momentum of an individual nucleon in the
nucleus-nucleus frame.  The number of charged hadrons produced in this 
tube-tube collision is the same as that produced in an elementary 
nucleon-nucleon collision with the same available energy (baryon masses 
subtracted).  That is, eq. (1) is applied with 
$M_{min} = (n_P+n_T)m_N+m_{\pi}$.  This means that knowledge of particle 
production in nucleon-nucleon collisions at energies much higher than 
200 GeV is required for RHIC!  There is no experimental information 
on nucleon-nucleon collisions above $\sqrt{s}$ = 62 GeV; higher energies 
should be measured in the future at RHIC.  There is data on proton-antiproton 
collisions from the UA5 collaboration at CERN \cite{pdg}.  The average 
multiplicity, exclusive of single diffractive events, 
may be represented by the function
\begin{equation}
N_{ch}^{p\bar{p}}(s) = 22 + 1.7 \ln\left(\sqrt{s}/200\right)
+ 5.1 \ln^2\left(\sqrt{s}/200\right)
\end{equation}
in the range $200 \leq \sqrt{s} \leq 900$ GeV.  (The best fit would give the 
multiplicity at 200 GeV as 21.4, not 22, but the latter number is chosen to 
match on continuously with the parametrization of eq. (1); it is still within 
the error bars of UA5.)  The CTM seems somewhat ambiguous when it comes to 
describing the rapidity distribution of the produced hadrons.  The most 
sensible approach is to assume it is a Gaussian peaked at the center-of-mass 
frame of the tube-tube system, that is, at $y_{cm}(n_P,n_T)$ determined by
\begin{equation}
\sqrt{s(n_P,n_T)}\sinh\left(y_{cm}(n_P,n_T)\right) = (n_P-n_T)p_{cm} \, .
\end{equation}
The dispersion of the Gaussian is assumed to be of the same form as in 
eq. (2), namely
\begin{equation}
D^2(n_P,n_T) = \ln\left(\frac{\sqrt{s(n_P,n_T)}}{(n_P+n_T)m_N}\right)
= D^2_{NN} + \thalf \ln\left(\frac{4n_Pn_T}{(n_P+n_T)^2}\right) \, .
\end{equation}
For collisions between equal size nuclei the last factor is on average zero.
As with the WNM the geometrical folding to compute the number of projectile 
and target nucleons is done the same way as with the LEXUS.

At this point it may be worth pointing out that the LEXUS, the WNM, and 
the CTM all reproduce nucleon-nucleon collisions by construction.  It is 
only the extrapolation to nucleus-nucleus collisions that is different.
Since the same nucleon-nucleon input and the same geometrical folding for 
nucleus-nucleus collisions is used in all three theories, any differences 
can only arise because of the different dynamics assumed as described above.
In particular there are no free parameters in any of these theories.  
One caveat is the numerical value 
of the inelastic nucleon-nucleon cross section $\sigma_{inel}$.  For the SPS 
energy range and below the total and elastic cross sections are relatively 
constant at 40 and 8 mb, respectively.  For the theories considered here it 
makes the most sense to exclude the single diffractive part of the inelastic 
cross section.  The LEXUS uses 24 mb, 
corresponding to hard inelastic collisions \cite{lexus}.  Above the SPS energy 
range the total and elastic cross sections rise to about 50 and 10 mb, 
respectively, at the full RHIC energy of $\sqrt{s}$ = 200 GeV.  In all of the 
calculations presented in this paper we use 30 mb for the inelastic cross 
section in the WNM and the CTM because that is what the original authors of 
those models used.  We have verified by direct calculation 
that the dependence of $dN_{ch}/dy$ on the value chosen for this cross section 
is negligible, typically a few percent.  

In Fig. 1 we plot the predictions for $dN_{ch}/dy$ from the three theories and 
the available data.  The first panel is for the 5\% most central Pb+Pb 
collisions at the SPS with $\sqrt{s}$ = 17 A GeV.  The data from NA49 contains 
identified electrically charged kaons and pions and is truly the rapidity 
density.  The next three panels are for the 6\% most central Au+Au collisions 
at RHIC with $\sqrt{s}$ = 56, 130 and 200 A GeV.  The data presented by 
PHOBOS does not identify the particles and is $dN_{ch}/d\eta$.  There is a 
Jacobian relating the rapidity and pseudo-rapidity distributions.  For a 
hadron of mass $m$ and momentum $p$ emerging at 90 degrees with respect to 
the beam, $y = \eta = 0$, the relationship is
\begin{equation}
\frac{dN_{ch}}{d\eta}(\eta = 0) = v \frac{dN_{ch}}{dy}(y = 0)
\end{equation}
where $v =p/\sqrt{p^2+m^2}$ is the velocity of that particle.  If most of the 
charged particles are pions with an average transverse momentum of 3 times 
their mass we have $v = 0.95$.  The two data points from PHOBOS plotted in 
Fig. 1 are 
the numbers quoted at the beginning of the paper multiplied by 1.05 to convert 
pseudo-rapidity to rapidity density.  This is a small effect and does not 
actually affect any conclusions, assuming that pions are indeed the most 
abundant charged hadrons.

The ordering of the three theories is easily understood.  The LEXUS produces 
more particles than the WNM because in the latter theory a nucleon, once 
wounded, cannot itself produce any more particles.  On the contrary, in the 
LEXUS a struck nucleon loses momentum but continues to produce particles on 
every subsequent collision.  The CTM produces fewer particles than the WNM as 
a consequence of the fact that particle production increases more slowly than 
$\sqrt{s}$.  For example, for a collision with $n_P=n_T\equiv n$ the ratio of 
particles produced by the WNM relative to the CTM is $\sqrt{n}$ if eq. (1) is 
used, and is even greater if eq. (5) comes into play.

The LEXUS represents the NA49 data very well, but predicts about 60\% more 
particles than is measured by PHOBOS at both energies.  The WNM represents the 
PHOBOS data reasonably well but predicts only about 70\% of the particles 
observed by NA49.  The CTM predicts far too few particles at all of these 
energies.  What interpretation can we give to these results?  One obvious 
possibility is that particle production at 17 A GeV is dominated by incoherent 
nucleon-nucleon interactions, but as the energy rises to 56 A GeV destructive 
interference plays an increasing role resulting in a reduction in the number 
of particles produced.  This is not the only interpretation one might give 
but it is the most obvious one in the context of these three theories. 
A caveat is that once produced, the particles (whether they initially be 
considered quarks and gluons or hadrons) can interact with one another and 
change the total particle number.  Generally one expects that the number of 
observed particles can be higher than originally produced, not lower, 
on account of the nondecrease 
of entropy.  Even if that occurs it cannot change our conclusion.  If extra 
particle production was invoked such as to make the WNM model agree with the 
NA49 data then it would predict too many particles compared to PHOBOS.

It will be very interesting to identify the charged hadrons measured by 
PHOBOS.  It will also be very interesting to discover whether any other 
observables display a similar change in going from SPS to RHIC energies.  
The HIJING model \cite{hijing} has already been compared to the PHOBOS data 
in their paper \cite{remark} and comparisons to other theories will surely 
follow.  With RHIC operational the game is afoot at last! 

\section*{Acknowledgements}

S.J. was supported by
the Director, Office of Energy Research, Office of High Energy and Nuclear 
Physics, Division of Nuclear Physics, and by the Office of Basic Energy 
Sciences, Division of Nuclear Sciences, of the U.S. Department of Energy
under Contract No. DE-AC03-76SF00098.  J.K. was supported by
the Department of Energy under grant DE-FG02-87ER40328.

%\newpage

\section*{Figure Captions}

\noindent
Figure 1: The charged hadron rapidity distributions for central Pb+Pb 
collisions (SPS) and Au+Au collisions (RHIC) at the indicated energies.
The SPS data is from NA49 \cite{na49} and the RHIC data is from PHOBOS 
\cite{phobos}.  The solid (top) curves are LEXUS, the dashed (middle) 
curves are the wounded nucleon 
model, and the dot-dashed (bottom) curves are the collective tube model.


\begin{thebibliography} {40}
\bibitem{remark} HIJING distributions were used by PHOBOS to correct 
their multiplicities.  Also, the comparison to HIJING predictions was 
phrased in terms of the number of participant pairs which itself was 
calculated from HIJING.  In our opinion this is not a totally unbiased 
comparison of theory and experiment.
\end{thebibliography}
\end{document}